\documentclass[aps,showpacs,preprintnumbers,amsmath,amssymb]{revtex4}
\usepackage{dcolumn}
\usepackage[dvips]{epsfig}

\usepackage{float}
\usepackage{graphics,epsfig}
\usepackage{graphicx}
\usepackage{dcolumn}
\usepackage{bm}

\begin{document}
\baselineskip=0.8 cm
\title{\bf Effects of Homogeneous Plasma on Strong Gravitational Lensing of Kerr Black Holes }

\author{
Changqing Liu$^{1}$,\footnote{Electronic address:
lcqliu2562@163.com}, Chikun Ding$^{1}$, \footnote{Electronic
address: : Chikun Ding@huhst.edu.cn}, and Jiliang
Jing$^{2}$\footnote{Electronic address:jljing@hunnu.edu.cn} }

\affiliation{1) Department of Physics, Hunan University of
Humanities Science and Technology, Loudi, Hunan 417000, P. R. China}

\affiliation{2) Department of Physics, and Key Laboratory of Low
Dimensional Quantum Structures and Quantum Control of Ministry of
Education, Hunan Normal University,  Changsha, Hunan 410081, P. R.
China}

\begin{abstract}
\baselineskip=0.6 cm
\begin{center}
{\bf Abstract}
\end{center}

Considering a Kerr black hole surrounded by a homogenous
unmagnetised plasma medium, we study the strong gravitational
lensing on the equatorial plane of the Kerr black hole. We find that
the presence of the uniform plasma can increase the photon-sphere
radius $r_{ps}$, the coefficient $\bar{a},\bar{b}$, the angular
position of the relativistic images $\theta_{\infty}$, the
deflection angle $\alpha(\theta)$ and the angular separation $s$.
However the relative magnitudes $r_m$ decreases in the presence of the
uniform plasma medium. It is also shown that the impact of the
uniform plasma on the effect of strong gravitational become smaller
as the spin of the Kerr black increace in prograde orbit($a>0$).
Especially, for the extreme black hole(a=0.5), the effect of strong
gravitational lensing in homogenous plasma medium is the same as the
case in vacuum for the prograde orbit.
\end{abstract}

 \maketitle
\newpage
\section{Introduction}

It is fact that a plasma exists around a supermassive black hole at
the centre of the Galaxy\cite{noh}. Consider a plasma cloud
surrounding a black hole, the light's propagation deviates from
lightlike geodesics in a way that depends on the frequency of the
plasma. Thus, compared with the general relativistic vacuum
propagation effects, the astrophysical plasma as refractive and
dispersive medium can influent the relativistic images due to the
gravitational lensing and the shape of accretion disk effect ranging
from pulsars and X-ray binaries to active galactic
nucleus\cite{Blandford,Bicak,Krikorian1985,Krikorian1999}. The study
of astrophysical processes in plasma medium surrounding black hole
becomes very interesting and important.

A general theory of geometrical optics in a curved space-time, in an
isotropic dispersive medium was  proposed in the textbook of
Synge\cite{Synge}. It is based on the elegant abstract Hamiltionian
thery of rays and waves. In the book of Perlick \cite{Perlick} the
method was developed for consideration of a deflection
of the light rays in presence of the gravity and plasma.
The general formulae for the exact light deflection angle
in the Schwarzschild and Kerr metric, in presence of
plasma with spherically symmetric distribution of concentration,
are obtained in the form of integrals. The generally covariant equations
describing the propagation of waves with an arbitrary dispersion
relation in a nonuniform, unmagnetised  plasma medium was derived in
paper\cite{Kulsrud}. The geometric optics approximation through a
magnetised plasma in the vicinity of a compact object was presented
in paper\cite{Blandford}. On the basis of his general approach of
the geometric optics. the gravitational lensing in inhomogeneous and
homogeneous plasma around black holes has been recently studied
in~\cite{BKTs2009,BKTs2010,TsBK2013,Moro2013,ErMao2013,Atamurotov,Rogers2015,our2015}
as extension of vacuum studies. However,their research work
restricted to the static spacetime and slowly rotating compacted
object in plasma medium.

In this paper, we will study the strong gravitational lensing by
Kerr black hole in a unmagnetised  homogeneous plasma medium. It is
well known that the gravitational lensing is regarded as a powerful
indicator of the physical nature of the central celestial objects.
It is shown that the relativistic images due to the gravitational
lensing effect carry some essential signatures about the central
celestial objects and could provide the profound verification of
alternative theories of gravity in their strong field regime
\cite{Darwin,Vir,Vir1,Vir3,
Bozza2,Bozza3,Gyulchev,Fritt,Bozza1,Eirc1,whisk,Bhad1,Song2,TSa1,
Ls1,Keeton}. The main purpose of this paper is to study the strong
gravitational lensing by the Kerr Black hole in the unmagnetised
homogeneous plasma medium and extend the results of the paper of
Bisnovatyi-Kogan and Tsupko \cite{BKTs2010} to the case of rotating
gravitational lens, and to see the impact of Homogeneous Plasma on
photon sphere radius, the deflection angle, the coefficients and the
observable quantities of strong gravitational lensing. Moreover, we
will explore how it differs from the Kerr black hole lensing in
vacuum.

The paper is organized as follows: In Sec. II, we will derive the
expression for the deflection angle of light in Kerr black hole in
the presence of homogeneous plasma. In Sec. III, we study the
physical properties of the strong gravitational lensing by Kerr
black hole and probe the effects of homogeneous plasma on the
deflection angle, the coefficients and the observable quantities for
gravitational lensing in the strong field limit. We end the paper
with a summary.

\section{Rotating Kerr spacetime and radius of photon sphere}
 Considering  a rotating Kerr black hole surrounded by plasma.
The Kerr metric in the standard Boyer-Lindquist coordinates can be
expressed as
\begin{eqnarray}\label{kerr}
ds^2=-\Big(1-\frac{2Mr}{\Sigma}\Big)~dt^2-
\Big(\frac{4Mar\sin^2\theta} {\Sigma}\Big)~dtd\phi
+\frac{\Sigma}{\Delta}~dr^2+\Sigma~d\theta^2+
\Big(r^2+a^2+\frac{2Ma^2r\sin^2\theta}{\Sigma}\Big)
\sin^2\theta~d\phi^2,
\end{eqnarray}
with
\begin{eqnarray}
\Delta\equiv r^2-2Mr+a^2,~~~
\Sigma\equiv r^2+a^2\cos^2~\theta.
\end{eqnarray}
We assume that the spacetime is filled with a non magnetized cold
plasma whose electron plasma frequency $w_p$ is a function of the
radius coordinate only,
\begin{equation}
\omega_p(r)^2 = \frac{4\pi e^2}{m} N(r).
\end{equation}
Here $e$ is the charge of the electron, $m$ is the electron mass,
and $N(r)$ is the number density of the electrons in the plasma.
When $\omega_p$ is constant, plasma is homogeneous.  In this paper
we only consider homogeneous plasma. The refraction index $n$ of
this plasma depends on the plasma frequency $\omega_p$ and on the
frequency $\omega$ of the photon as it is measured by a static
observer,
\begin{equation}
n ^2 = 1 - \frac{\omega_p^2}{\omega^2}.
\end{equation}
Let us now study the strong gravitational lensing of the rotating
Kerr black hole surrounded by plasma. As in refs.
\cite{Bozza2,Bozza3,Gyulchev,Fritt,Bozza1,Eirc1,whisk,Bhad1,Song2,TSa1,
Ls1}, we just consider that both the observer and the source lie in
the equatorial plane of the black hole and the whole trajectory of
the photon is limited on the same plane. Using the condition
$\theta=\pi/2$ and taking $2M=1$, the metric (\ref{kerr}) is reduced
to
\begin{eqnarray}
ds^2&=&-A(r)dt^2+B(r)dr^2+C(r) d\phi^2-2D(r)dtd\phi, \label{metric1}
\end{eqnarray}
with
\begin{eqnarray}
A(r)&=&1-\frac{1}{r},\\
B(r)&=&\frac{r^2}{a^2-r+r^2},\\
C(r)&=&a^2+\frac{a^2}{r}+r^2,\\
D(r)&=&\frac{a}{r}.
\end{eqnarray}
The general relativistic geometrical optics on the background of the
curved space-time, in a refractive and dispersive plasma medium, was
developed by Synge \cite{Synge}. Based on the Hamiltonian approach
for the description of the geometrical optics. The Hamiltonian for
the photon around the Kerr black hole surrounded by plasma has the
following form\cite{Kulsrud}

\begin{equation}
H(x^i,p_i) = \frac{1}{2} \left[ g^{ik} p_i p_k +\hbar^2 \omega_p^2
\right] = 0 \, . \label{H-definition}
\end{equation}
It is interested to notice that the expression of the Hamiltonian
for the photon around the Kerr black hole surrounded by homogeneous
plasma is similar to the Hamiltonian of the massive particle in
vacuum.
 Using the Hamiltonian for the photon around
the Kerr black hole, we can get the Hamiltonian differential
equations.
\begin{equation}
\label{D-Eq} \frac{dx^i}{d \lambda} = \frac{\partial H}{\partial
p_i}  \, , \; \; \frac{dp_i}{d \lambda} = - \frac{\partial
H}{\partial x^i} \, ,
\end{equation}
then we get two costants of motions are the energy and the angular
momentum of the particle
\begin{equation}
E=-p_t=\hbar\omega,~~~ L=P_\phi.
\end{equation}
Let us consider a homogeneous plasma with $\omega _p= const$. We
introduce notations of $\hat{E}$ and $\hat{L}$
\begin{equation} \label{definition-E-L}
\frac{-p_t}{\hbar \omega_p} = \frac{\omega}{\omega_p} = \hat{E} > 1
\, , \quad \frac{L}{\hbar \omega_p} =\hat{ L} > 0.
\end{equation}
 From this equations, we find an expression for the
 $\dot{t},\dot{r},\dot{\phi}$ in terms of $\hat{E}$ and $\hat{L}$
\begin{eqnarray}
\frac{dt}{d\lambda}&=&\frac{\hbar \omega_p(\hat{E}C(r)-\hat{L}D(r))}{D(r)^2+A(r)C(r)},\label{u3}\\
\frac{d\phi}{d\lambda}&=&\frac{\hbar \omega_p(\hat{E}D(r)+\hat{L}A(r))}{D(r)^2+A(r)C(r)},\label{u4}\\
\bigg(\frac{dr}{d\lambda}\bigg)^2&=&\frac{\hbar^2
\omega_p^2(\hat{E^2}C(r)-(A(r)C(r)+D(r)^2)-2\hat{E}\hat{L}D(r)-\hat{L}^2A(r))}{(B(r)[D(r)^2+A(r)C(r)]}.
\end{eqnarray}
Where $\lambda$ is an affine parameter along the geodesics. With the
condition $\frac{dr}{d\lambda}|_{r=r_0}=0 $, we can obtain the
angular momentum $\hat{L}(r_0)$
\begin{eqnarray}
\hat{L}(r_0)=\frac{-\hat{E}D(r_0)+\sqrt{A(r_0)PC(r_0)+\hat{E}^2D^2(r_0)}}{A(r_0)},\\
PC(r_0)=\hat{E^2}C(r_0)-(A(r_0)C(r_0)+D(r_0)^2).
\end{eqnarray}
For the photon moving in the plasma have the effective
mass\cite{BKTs2010,Kulsrud} $m_{eff}=\hbar \omega_p$, the impact
parameter can be expressed as
\begin{eqnarray}
u=\frac{L}{\sqrt{E^2-m_{eff}^2}}=\frac{\hat{L}}{\sqrt{\hat{E}^2-1}}.
\end{eqnarray}
 Moreover, the photon sphere is a time-like hyper-surface
($r=r_{ps}$) on which the deflect angle of the light becomes
unboundedly large as $r_0$ tends to $r_{ps}$. In this spacetime, the
equation for the photon sphere reads
\begin{eqnarray}\label{rpsss}
A(r)PC'(r)-A'(r)PC(r)+2\hat{L}\hat{E}[A'(r)D(r)-A(r)D'(r)]=0.\label{root0}
\end{eqnarray}
The biggest real root external to the horizon of this equation is
defined as the radius of the photon sphere $r_{ps}$. In the case of
the Kerr black hole surrounded by the plasma, the analytical
expression of marginally circular photon orbits takes a form
\begin{eqnarray}\label{rpsssb}
a^4
&+&2a^2(2-a^2)r+(4+8a^2+a^4-4\hat{E}^2a^2-4\hat{E}^4a^4+2a^2(4-3\hat{E}^2)-
4a^2\hat{E}^2(\hat{E}^2-1))r^2\nonumber\\&+&(-8-4a^2+4a^2\hat{E}^2
-4(4-3\hat{E}^2)(1+a^2-a^2\hat{E}^2)+4a^2(\hat{E}^4-1))r^3
\nonumber\\&+&(4+(4-3\hat{E}^2)(12+2a^2-7a^2\hat{E}^2)+4(\hat{E}^2-1)(a^2\hat{E}^2-2a^2-2))r^4
\nonumber\\&+&((4-3\hat{E}^2)(6\hat{E}^2-16)+4(\hat{E}^2-1)(8+a^2
-2a^2\hat{E}^2-3\hat{E}^2))r^5\nonumber\\&+&((4-3\hat{E}^2)^2+4(\hat{E}^2-1)(5\hat{E}^2-11))r^6+
4(\hat{E}^2-1)(6-5\hat{E}^2)r^7+4(\hat{E}^2-1)^2r^8=0.
\end{eqnarray}

\begin{figure}[ht]
\begin{center}
\includegraphics[scale=1.1]{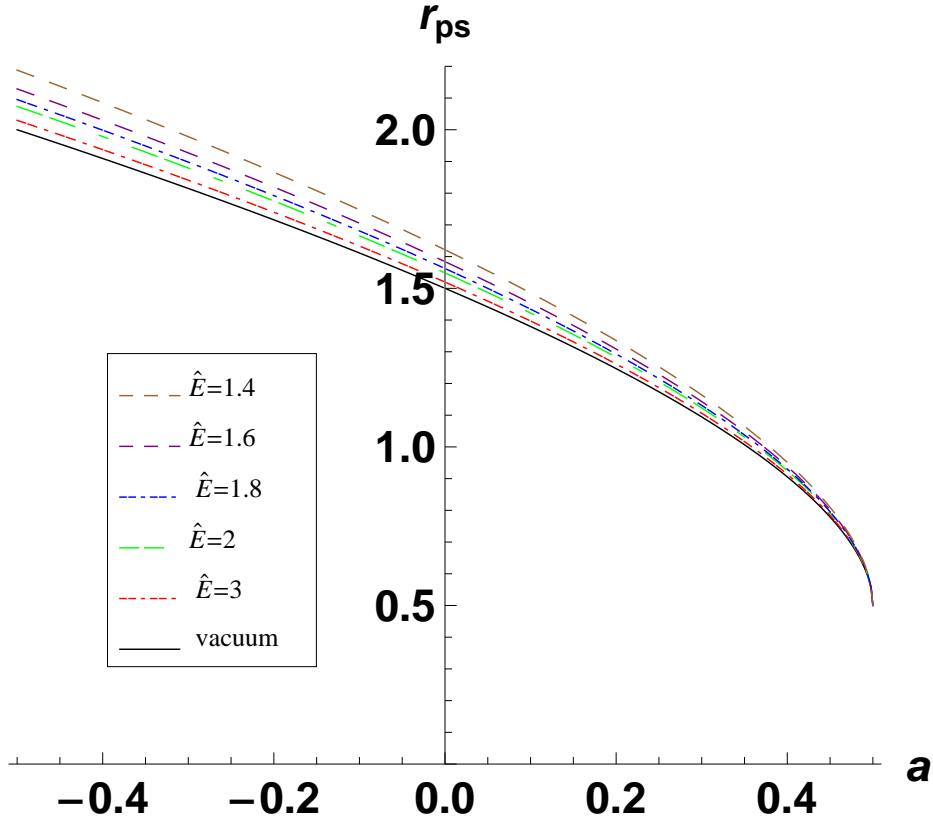}
\caption{Variation of the radius of the photon sphere  $r_{ps}$ with
the parameter $a$ in different plasma medium and vacuum of the Kerr
black hole. parameter $\hat{E}=\frac{\omega}{{\omega_p}}$ describe
the ratio photon frequency to the plasma frequency. \label{figure1}}
\end{center}
\end{figure}

Obviously, circular photon orbits is depend on the plasma frequency.
 Especially, for the photon radius in a
static Schwarzschild black hole surrounded by homogeneous plasma
have the analytical expression from the equation (\ref{rpsss})
 \begin{eqnarray}\label{rbsss}
 r_{ps}=\frac{3\hat{E}^2-4+\hat{E}\sqrt{-8+9\hat{E}^2}}{4(\hat{E}^2-1)}.
\end{eqnarray}
This just the result obtained in reference\cite{BKTs2010}. As
$\hat{E}\rightarrow\infty$,  the photon radius
$r_{ps}\rightarrow\frac{3}{2}$ corresponds to the photons of the
static Schwarzschild black hole in the vacuum. In
Fig.(\ref{figure1}), we present the variation of the photon-sphere
radius $r_{ps}$ with the rotational parameter $a$ in different
plasma medium and vacuum. as expected, compared with the case in the
vacuum, the presence of plasma can increase the photon sphere radius
$r_{ps}$. It also shown that the growth of the photon-sphere radius
in prograde orbit($a>0$) is less than the case in retrograde
orbit($a<0$). Especially, for the
 extreme black hole(a=0.5), the photon sphere radius in plasma medium
 is the same as in  vacuum for the  prograde orbit.

\section{Strong gravitational lensing of rotating Kerr black hole in plasma medium}

In this section we will study the gravitational lensing of the
rotating Kerr black hole in plasma medium which has a photon sphere
and then probe the impact of plasma on the coefficients and the
observable quantities of the strong gravitational lensing.

\subsection{Coefficients of strong gravitational lensing}

The deflection angle for the photon coming from infinite in a
stationary, Kerr black hole in plasma medium can be given as follow:
\begin{eqnarray}
\alpha(r_{0})=I(r_{0})-\pi,
\end{eqnarray}
with
\begin{eqnarray}
I(r_0)=2\int^{\infty}_{r_0}\frac{\sqrt{B(r)|A(r_0)|}[\hat{E}D(r)+\hat{L}A(r)]dr}{\sqrt{D^2(r)+A(r)C(r)}
\sqrt{sgn(A(r_0))[A(r_0)PC(r)-A(r)PC(r_0)+2\hat{E}\hat{L}[A(r)D(r_0)-A(r_0)D(r)]]}},\nonumber\\\label{int1}
\end{eqnarray}
where $sgn(X)$ gives the sign of $X$.

It is obvious that the deflection angle increases as the parameter
$r_0$ decreases. For a certain value of $r_0$ the deflection angle
becomes $2\pi$, so that the light ray makes a complete loop around
the lens before reaching the observer. If $r_0$ is equal to the
radius of the photon sphere $r_{ps}$, we can find that the
deflection angle diverges and the photon is captured by the compact object.

In order to find the behavior of the deflection angle when the photon is close to the photon sphere, we
 use the evaluation method proposed by Bozza \cite{Bozza2}. The divergent
integral in Eq. (\ref{int1}) is first split into the divergent part
$I_D(r_0)$ and the regular one $I_R(r_0)$, and then both of them are
expanded around $r_0=r_{ps}$ with sufficient accuracy. This
technique has been widely used in the study of the strong
gravitational lensing for various black holes
\cite{Bozza2,Bozza3,Gyulchev,Fritt,Bozza1,Eirc1,whisk,Bhad1,Song2,TSa1,
Ls1}. Let us now to define a variable
\begin{figure}[ht]
\begin{center}
\includegraphics[scale=0.73]{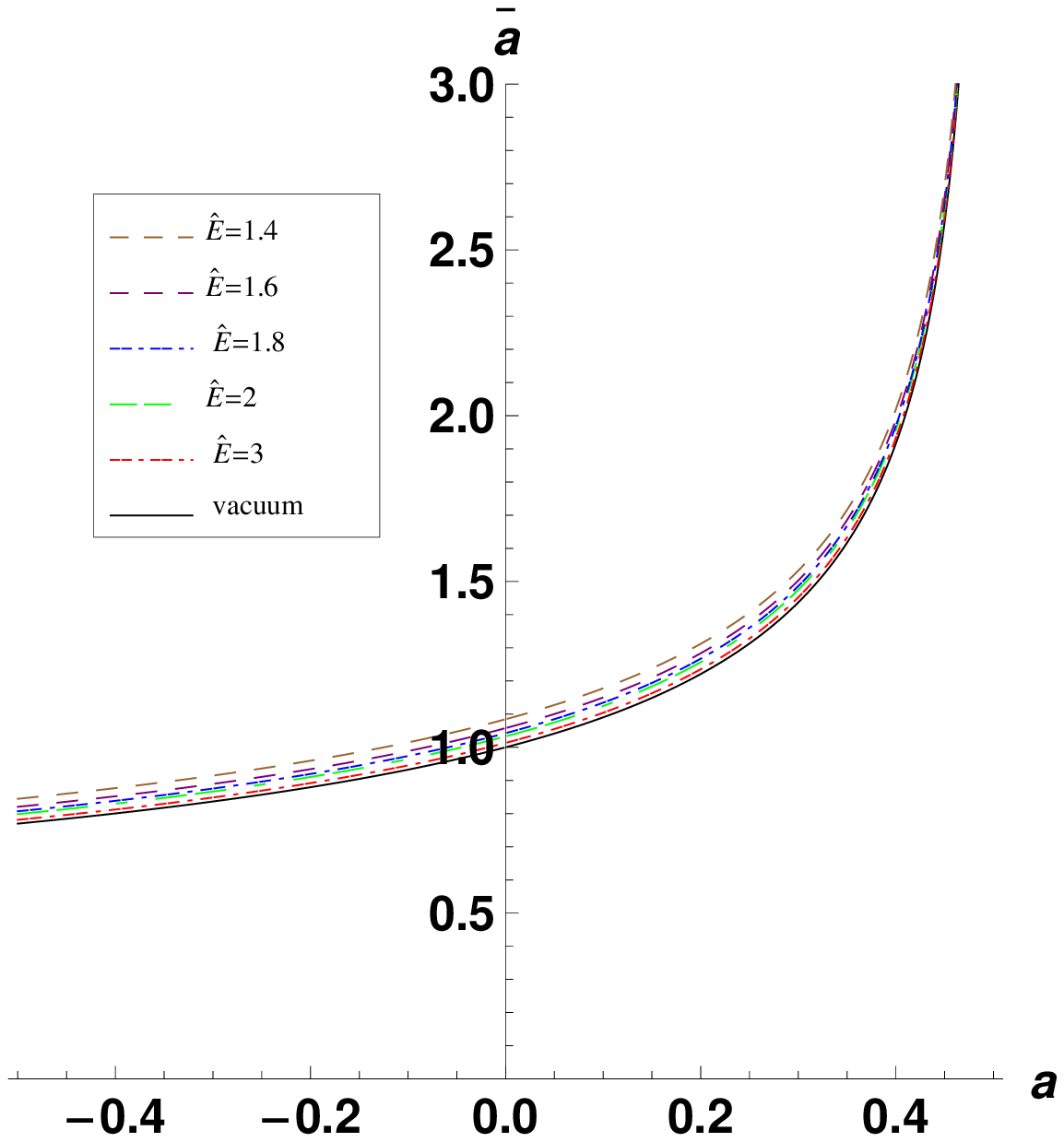},\includegraphics[scale=0.73]{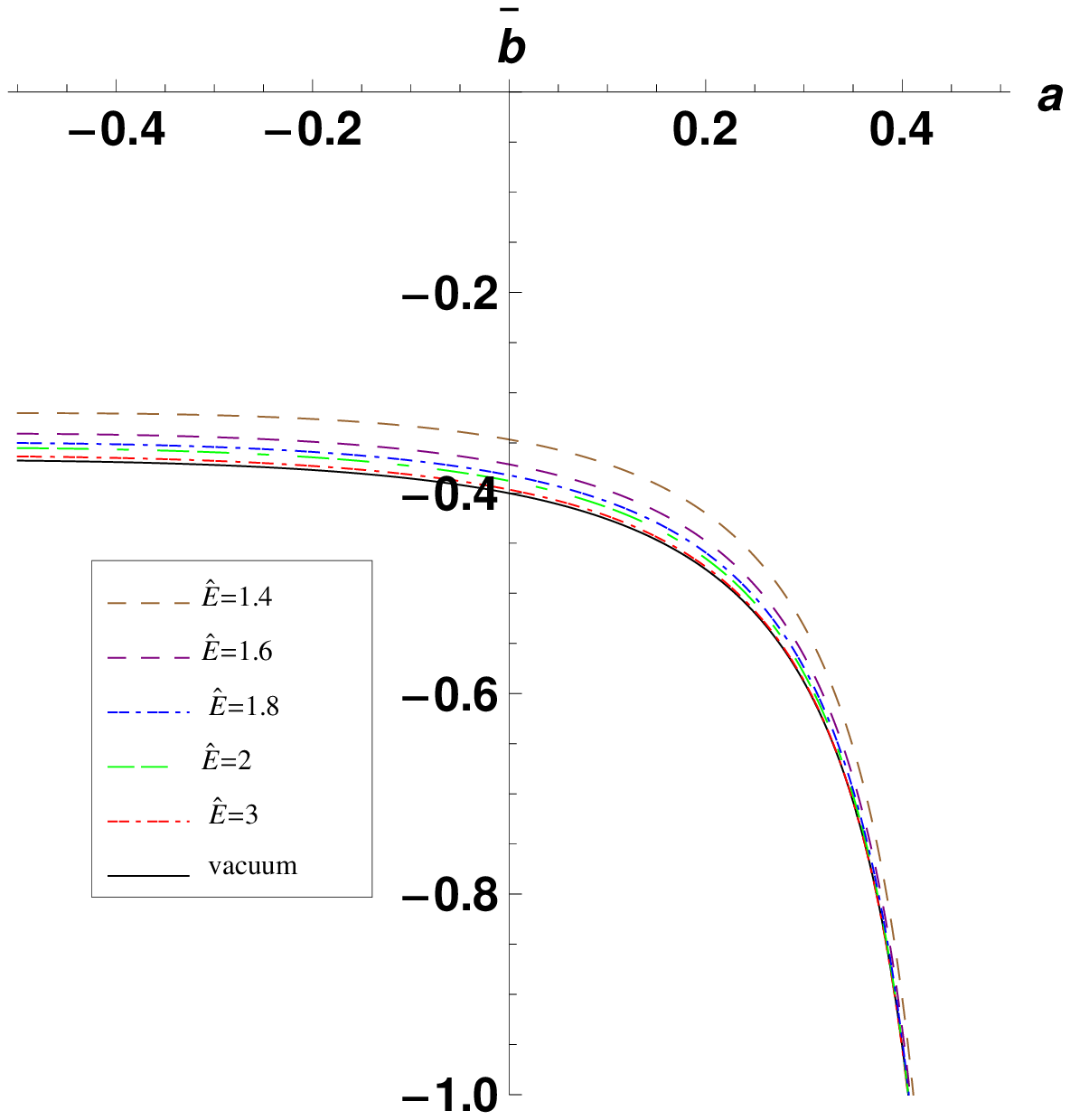}\caption{Variation of the coefficients $\bar{a}$, $\bar{b}$ for the
strong gravitational lensing  with  the parameter $a$ in different
plasma medium and vacuum of the Kerr black hole. parameter
$\hat{E}=\frac{\omega}{{\omega_p}}$ describe the ratio of photon
frequency to the plasma frequency.} \label{ coefficients}
\end{center}
\end{figure}
\begin{figure}[ht]
\begin{center}
\includegraphics[scale=1.0]{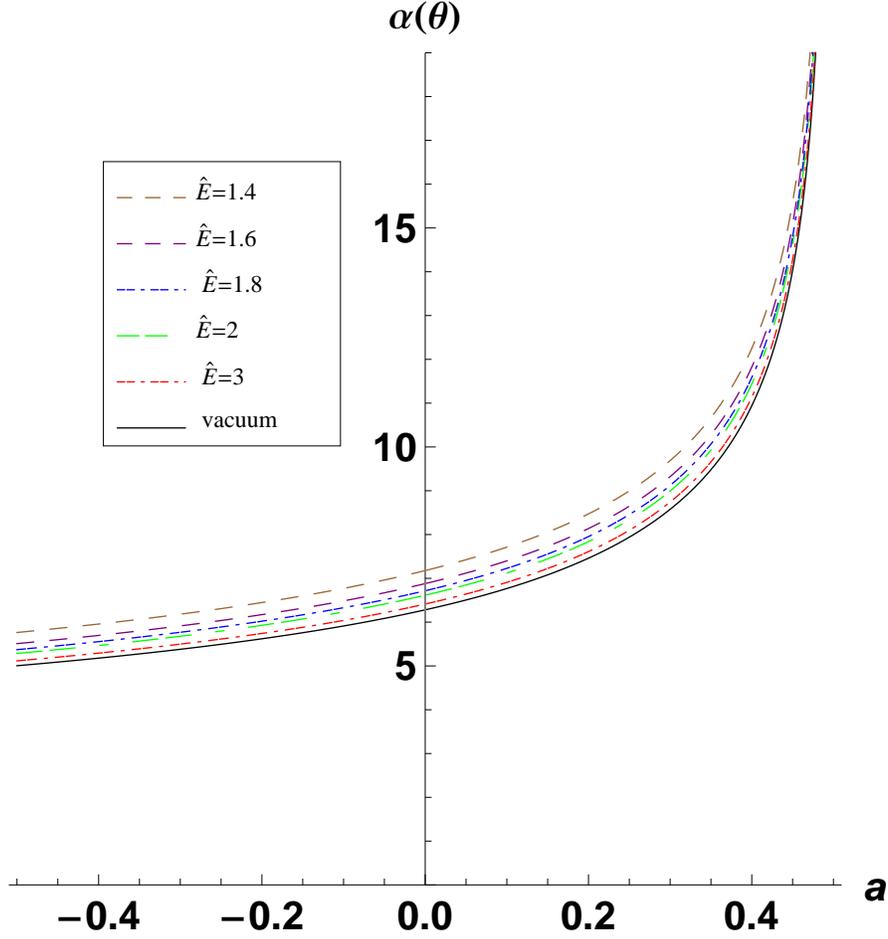}\caption{Variation of deflection angles  $\alpha(\theta)$ evaluated at $u=u_{ps}+0.00326$ with  with  the
parameter $a$ in different plasma medium and vacuum of the Kerr
black hole. parameter $\hat{E}=\frac{\omega}{{\omega_p}}$ describe
the ratio of photon frequency to the plasma
frequency.}\label{figure3}
\end{center}
\end{figure}
\begin{eqnarray}
z=1-\frac{r_0}{r},
\end{eqnarray}
and rewrite the Eq.(\ref{int1}) as
\begin{eqnarray}
I(r_0)=\int^{1}_{0}R(z,r_0)f(z,r_0)dz,\label{in1}
\end{eqnarray}
with
\begin{eqnarray}
R(z,r_0)&=&\frac{2r_0}{\sqrt{PC(z)}(1-z)^2}\frac{\sqrt{B(z)|A(r_0)|}[\hat{E}D(z)+\hat{L}A(z)]}{\sqrt{D^2(z)+A(z)C(z)}},
\end{eqnarray}
\begin{eqnarray}
f(z,r_0)&=&\frac{1}{\sqrt{sign(A(r_0))[A(r_0)-A(z)\frac{PC(r_0)}{PC(z)}+\frac{2\hat{E}\hat{L}}{PC(z)}(A(z)D(r_0)-A(r_0)D(z))]}}.
\end{eqnarray}
Obviously, the function $R(z, r_0)$ is regular for all values of $z$
and $r_0$. However, the function $f(z, r_0)$ diverges as $z$ tends
to zero, i.e., as the photon approaches the photon sphere. Thus, the
integral (\ref{in1}) can be separated into two parts $I_D(r_0)$ and
$I_R(r_0)$
\begin{eqnarray}
I_D(r_0)&=&\int^{1}_{0}R(0,r_{ps})f_0(z,r_0)dz, \nonumber\\
I_R(r_0)&=&\int^{1}_{0}[R(z,r_0)f(z,r_0)-R(0,r_0)f_0(z,r_0)]dz
\label{intbr}.
\end{eqnarray}
Expanding the argument of the square root in $f(z,r_0)$ to the
second order in $z$, we have
\begin{eqnarray}
f_0(z,r_0)=\frac{1}{\sqrt{p(r_0)z+q(r_0)z^2}},
\end{eqnarray}
where
\begin{eqnarray}
p(r_0)&=&\frac{r_0}{PC(r_0)}\bigg\{A(r_0)PC'(r_0)-A'(r_0)PC(r_0)+2\hat{E}\hat{L}[A'(r_0)D(r_0)-A(r_0)D'(r_0)]\bigg\},  \nonumber\\
q(r_0)&=&\frac{r_0}{2PC^2(r_0)}\bigg\{2\bigg(PC(r_0)-r_0PC'(r_0)\bigg)\bigg([A(r_0)PC'(r_0)-A'(r_0)PC(r_0)]
+2\hat{E}\hat{L}[A'(r_0)D(r_0)-A(r_0)D'(r_0)]\bigg)\nonumber\\&&+r_0PC(r_0)
\bigg([A(r_0)PC''(r_0)-A''(r_0)PC(r_0)]+2\hat{E}\hat{L}[A''(r_0)D(r_0)-A(r_0)D''(r_0)]\bigg)\bigg\}.\label{al0}
\end{eqnarray}
From Eq.(\ref{al0}), we can find that if
$r_{0}$ approaches the radius of photon sphere $r_{ps}$ the
coefficient $p(r_{0})$ vanishes and the leading term of the
divergence in $f_0(z,r_{0})$ is $z^{-1}$, which implies that the
integral (\ref{in1}) diverges logarithmically. The coefficient
$q(r_0)$ takes the form
\begin{eqnarray}
q(r_{ps})&=&\frac{sgn(A(r_{ps}))
r^2_{ps}}{2PC(r_{ps})}\bigg\{A(r_{ps})PC''(r_{ps})-A''(r_{ps})PC(r_{ps})+
2\hat{E}\hat{L}[A''(r_{ps})D(r_{ps})-A(r_{ps})D''(r_{ps})]\bigg\}.\nonumber\\
\end{eqnarray}
Therefore, the deflection angle in the strong field region can be
expressed as \cite{Bozza2}
\begin{eqnarray}
\alpha(\theta)=-\bar{a}\log{\bigg(\frac{\theta
D_{OL}}{u_{ps}}-1\bigg)}+\bar{b}+\mathcal{O}(u-u_{ps}), \label{alf1}
\end{eqnarray}
with
\begin{eqnarray}
&\bar{a}&=\frac{R(0,r_{ps})}{2\sqrt{q(r_{ps})}}, \nonumber\\
&\bar{b}&= -\pi+b_R+\bar{a}\log{\bigg\{\frac{2q(r_{ps})PC(r_{ps})}{u_{ps}\sqrt{\hat{E}^2-1}|A(r_{ps})
|[\hat{E}D(r_{ps})+\hat{L}_{ps}A(r_{ps})]}\bigg\}}, \nonumber\\
&b_R&=I_R(r_{ps}), \nonumber\\
&u_{ps}&=\frac{-\hat{E}D(r_{ps})+\sqrt{A(r_{ps})PC(r_{ps})+\hat{E}^2D^2(r_{ps})}}{A(r_{ps})\sqrt{\hat{E}^2-1}},\label{coa1}
\end{eqnarray}
where the quantity $D_{OL}$ is the distance between observer and
gravitational lens. Making use of Eqs. (\ref{alf1}) and
(\ref{coa1}), we can study the properties of strong gravitational
lensing in the rotating Kerr black hole in presence of homogenous
plasma. In Fig. (\ref{ coefficients}), we plot the changes of the
coefficients $\bar{a}$ and $\bar{b}$ with $a$ for a different ratio
of photon frequency to plasma frequency $\omega_p$ . It is shown
that the coefficients ($\bar{a}$ and $\bar{b}$ ) in the strong field
limit are functions of the parameters $a$ and $\hat{E}$. Compared
with the vacuum the case, the presence of plasma can increase  the
coefficients $\bar{a}$ and $\bar{b}$. It is also shown that the
growth of the coefficients $\bar{a}$ and $\bar{b}$ in prograde
orbit($a>0$) is less than the case in retrograde orbit($a<0$).
Especially, for the extreme black hole(a=0.5), the coefficients
$\bar{a}$ and $\bar{b}$  in plasma medium is the same as in  vacuum
for the  prograde orbit. With the help of the coefficients $\bar{a}$
and $\bar{b}$,  we plot the change of the deflection angles
evaluated at $u=u_{ps}+0.00326$ with the rotational parameter $a$
for a different ratio of photon frequency to plasma frequency
$\omega_p$ in Fig. (\ref{figure3}). It is shown that in the strong
field limit the deflection angles  $\alpha(\theta)$
 have the similar properties of the
coefficient $\bar{a}$.

\subsection{Observable quantities of strong gravitational lensing}

Let us now to study the effect of the homogeneous plasma medium on
the observable quantities of strong gravitational lensing. Here we
consider only the case in which the source, lens and observer are
highly aligned so that the lens equation in strong gravitational
lensing can be approximated well as \cite{Bozza3}
\begin{eqnarray}
\gamma=\frac{D_{OL}+D_{LS}}{D_{LS}}\theta-\alpha(\theta) \; mod
\;2\pi,
\end{eqnarray}
where $D_{LS}$ is the lens-source distance and $D_{OL}$ is the
observer-lens distance, $\gamma$ is the angle between the direction
\begin{figure}[ht]
\begin{center}
\includegraphics[scale=0.75]{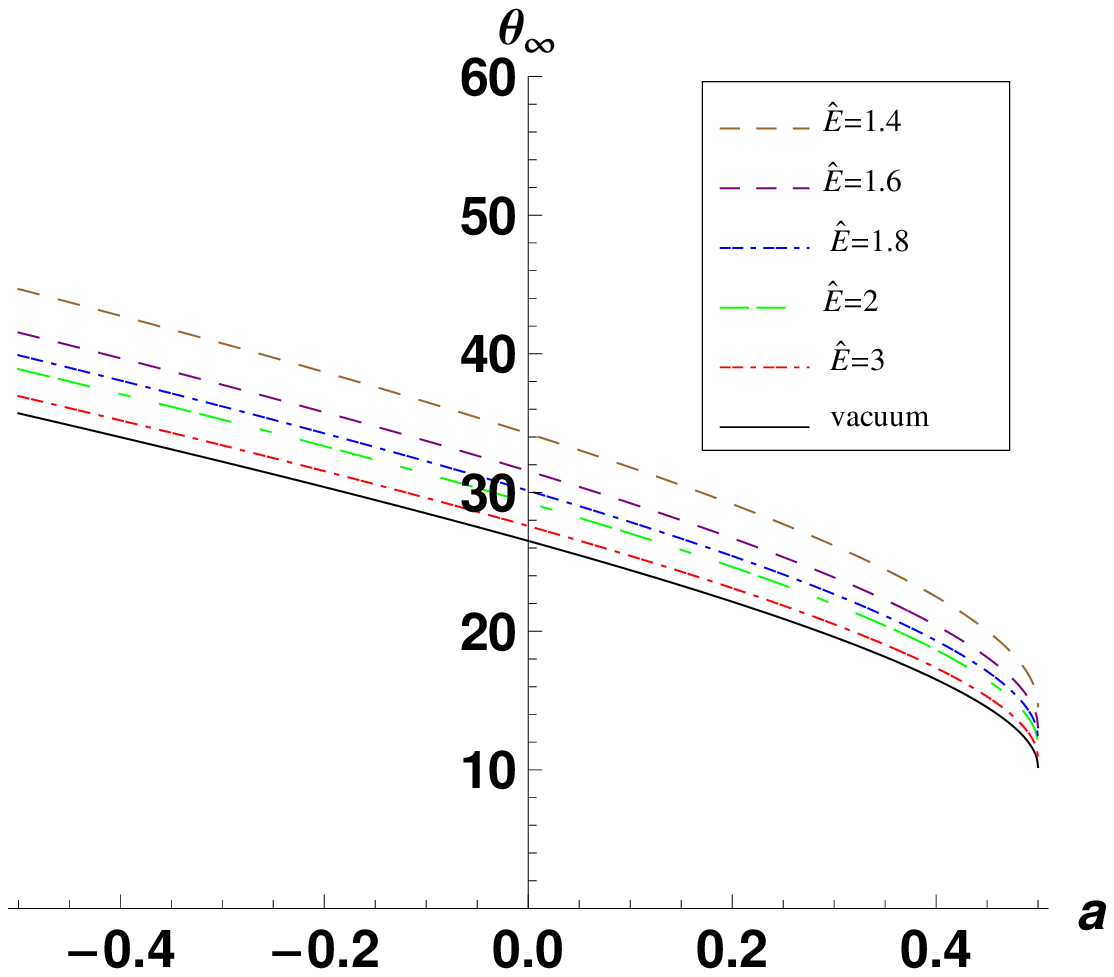},\includegraphics[scale=0.8]{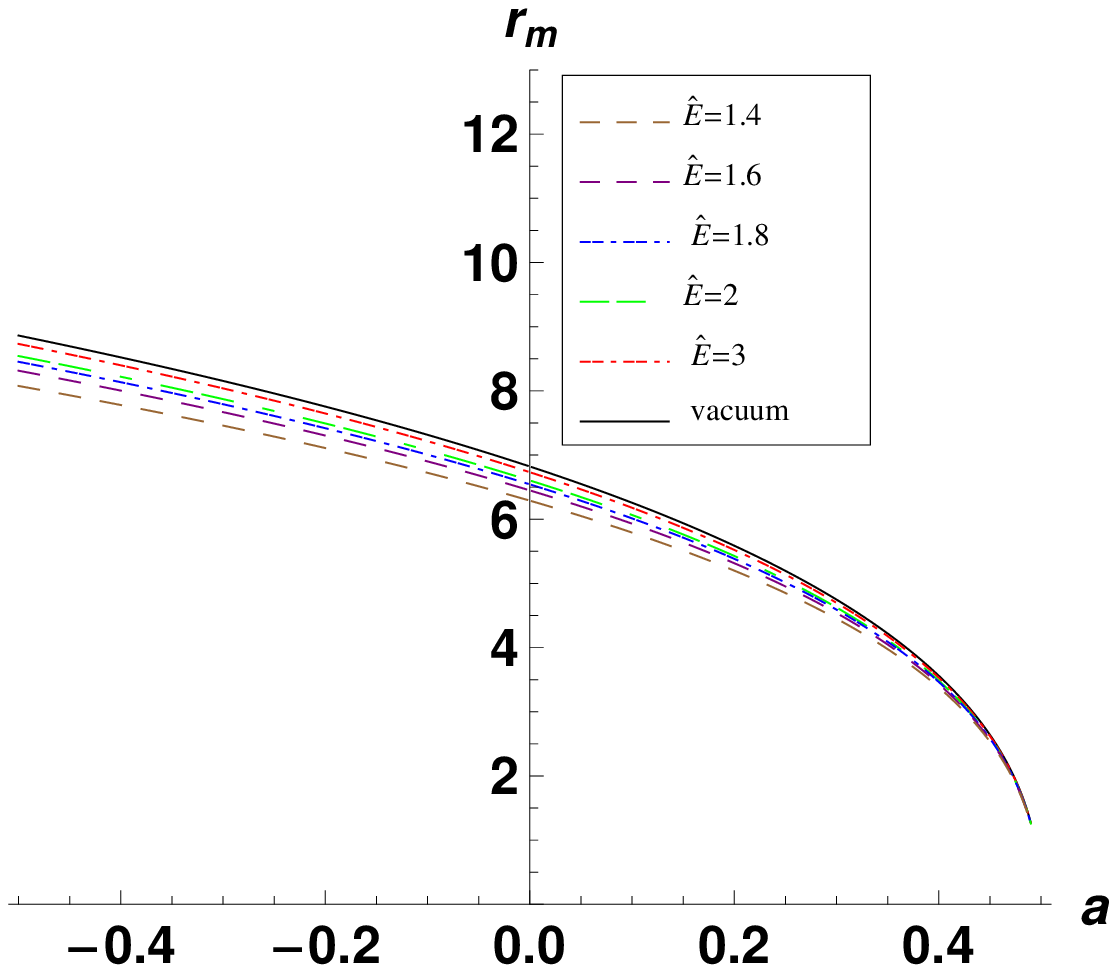},
\includegraphics[scale=0.8]{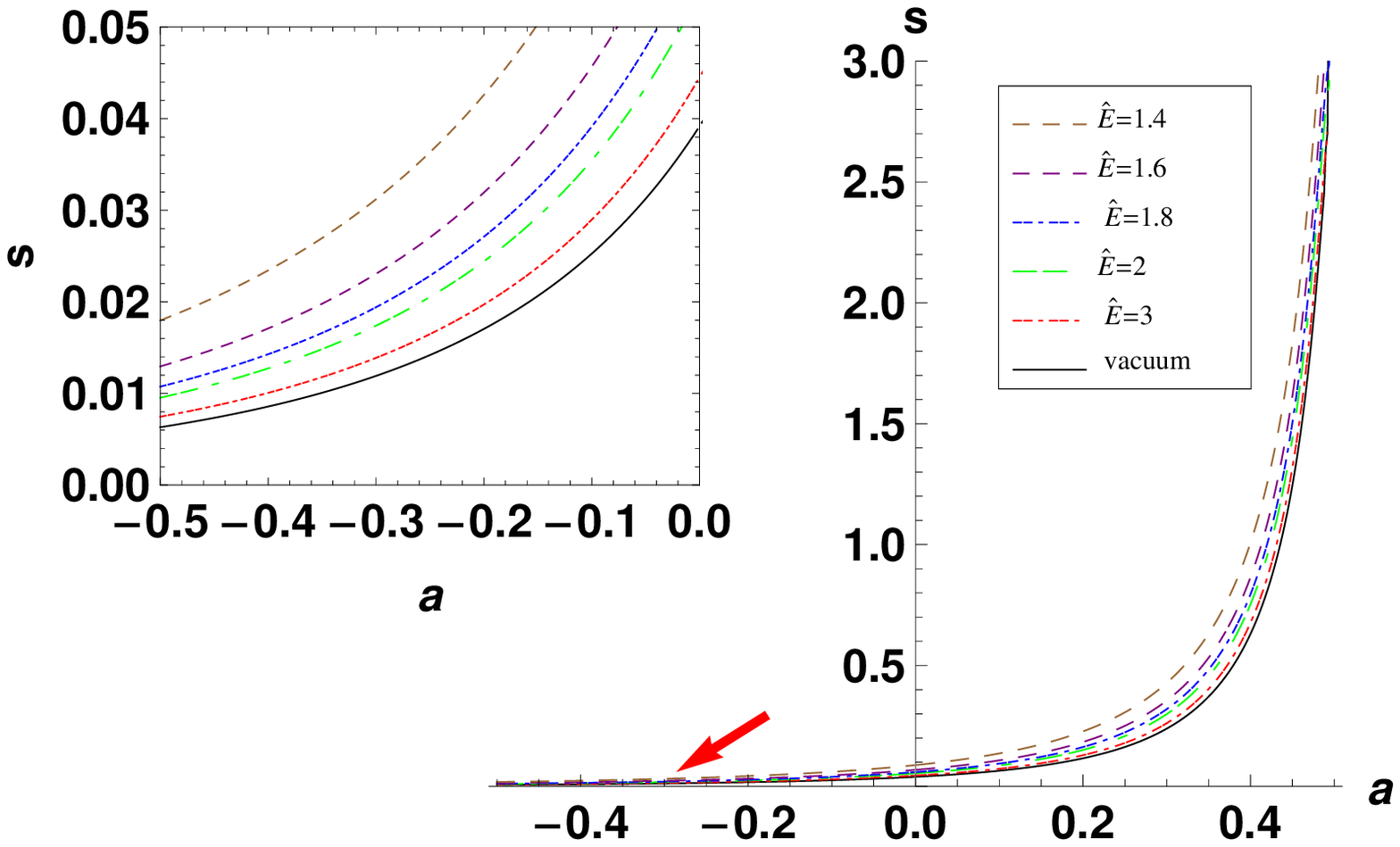}
\caption{Variation of the innermost relativistic image
$\theta_{\infty}$, the relative magnitudes $r_m$  and the angular
separation $s$  with  the parameter $a$ in different plasma medium
and vacuum of the Kerr black hole. parameter
$\hat{E}=\frac{\omega}{{\omega_p}}$ describe the ratio of photon
frequency to the plasma frequency.}\label{figure4}
\end{center}
\end{figure}
of the source and the optical axis, $\theta=u/D_{OL}$ is the angular
separation between the lens and the image. Following
ref. \cite{Bozza3}, we can find that the angular separation between
the lens and the n-th relativistic image is
\begin{eqnarray}
\theta_n\simeq\theta^0_n\bigg(1-\frac{u_{ps}e_n(D_{OL}+D_{LS})}{\bar{a}D_{OL}D_{LS}}\bigg),
\end{eqnarray}
with
\begin{eqnarray}
\theta^0_n=\frac{u_{ps}}{D_{OL}}(1+e_n),\;\;\;\;\;\;e_{n}=e^{\frac{\bar{b}+|\gamma|-2\pi
n}{\bar{a}}},
\end{eqnarray}
where the quantity $\theta^0_n$ is the image positions corresponding to
$\alpha=2n\pi$, and $n$ is an integer. According to the past
oriented light ray which starts from the observer and finishes at
the source the resulting images stand on the eastern side of the
black hole for direct photons ($a>0$) and are described by positive
$\gamma$. Retrograde photons ($a<0$) have images on the western side
of the compact object and are described by negative values of $\gamma$.
In the limit $n\rightarrow \infty$, we can find that
$e_n\rightarrow 0$, which means that the relation between the
minimum impact parameter $u_{ps}$ and the asymptotic position of a
set of images $\theta_{\infty}$ can be simplified further as
\begin{eqnarray}
u_{ps}=D_{OL}\theta_{\infty}.\label{uhs1}
\end{eqnarray}
In order to obtain the coefficients $\bar{a}$ and $\bar{b}$, we
needs to separate at least the outermost image from all the others.
As in refs. \cite{Bozza2,Bozza3},  we consider here the simplest
case in which only the outermost image $\theta_1$ is resolved as a
single image and all the remaining ones are packed together at
$\theta_{\infty}$. Thus the angular separation between the first
image and other ones can be expressed as \cite{Bozza2,Bozza3}
\begin{eqnarray}
s=\theta_1-\theta_{\infty}=\theta_{\infty} e^{\frac{\bar{b}-2\pi}{\bar{a}}}.\label{ss1}
\end{eqnarray}
By measuring $s$ and $\theta_{\infty}$, we can obtain the strong
deflection limit coefficients $\bar{a}$, $\bar{b}$ and the minimum
impact parameter $u_{ps}$. Comparing their values with those
predicted by the theoretical models, we can obtain information of
Kerr Black hole,

The mass of the central object of our Galaxy is estimated recently
to be $4.4\times 10^6M_{\odot}$ \cite{Genzel1} and its distance is
around $8.5kpc$, so that the ratio of the mass to the distance
$M/D_{OL} \approx2.4734\times10^{-11}$.  Making use of Eqs.
(\ref{coa1}), (\ref{uhs1}) and  (\ref{ss1})  we can estimate the
values of the coefficients and observable quantities for
gravitational lens in the strong field limit. The numerical value
for the angular position of the relativistic images
$\theta_{\infty}$, the angular separation $s$ and the relative
magnitudes $r_m$ are plotted in Fig. (\ref{figure4}). we find that
the variation of the angular position of the relativistic images
$\theta_{\infty}$ with the rotational parameter $a$ in different
plasma medium and vacuum is similar to that of the photon-sphere
radius $r_{ps}$.  However, the variation of the relative magnitudes
$r_m$ is contrary to the case of the photon-sphere radius $r_{ps}$.
We also find that the variation of the angular separation $s$ with
the rotational parameter $a$ in different plasma medium and vacuum
is similar to that of the deflection angle $\alpha(\theta)$.

\section{summary}

In this paper, we have investigated the strong gravitational lensing
of Kerr black hole surrounded by homegeneous plasma. We derived the
expression for the deflection angle of light in Kerr black hole in
the presence of homogeneous plasma and numerically calculated the
coefficient of the deflection angle. It is shown that the presence
of the uniform plasma increases the photon-sphere radius $r_{ps}$,
the coefficient $\bar{a},\bar{b}$, the angular position of the
relativistic images $\theta_{\infty}$, the deflection angle
$\alpha(\theta)$ and the angular separation $s$. However the
relative magnitudes $r_m$ decrease in presence of the uniform plasma
medium. It is also shown that the impact of the uniform plasma on
the effect of strong gravitational become smaller as the spin of
Kerr black increase in prograde orbit($a>0$). Especially, for the
extreme black hole(a=0.5), the effect of strong gravitational
lensing in homogenous plasma medium is the same as the case in
vacuum for the prograde orbit. In reality the plasma in the
neighborhood of the compact objects can be significantly
non-homogeneous. Such cases can be calculated numerically and much
more complicated, in the later research work we will consider the
influence of the non-homogeneous plasma medium on the strong
gravitational lensing of Kerr black hole.

\section{\bf Acknowledgments}

Changqing's work was supported by the National Natural Science
Foundation of China under Grant Nos.11447168. Chikun's work was
supported by the National Natural Science Foundation of China under
Grant Nos11247013; Hunan Provincial Natural Science Foundation of
China under Grant Nos. 12JJ4007 and 2015JJ2085.

\end{document}